\documentstyle[prl,aps]{revtex}
\begin{document} 

\title{Statistical mechanics of screened  
spatially indirect excitons}
 
\author{V.V. Nikolaev\footnote{Corresponding author;
E-mail address: V.Nikolaev@exeter.ac.uk} 
and M.E. Portnoi}
\address{
School of Physics, University of Exeter, Stocker Road, Exeter, EX4 4QL, UK\\
and A.F. Ioffe Physico-Technical Institute, St.-Petersburg, Russia}

\maketitle

\begin{abstract}
\noindent
We study thermodynamic properties of spatially separated 
electron-hole plasma in double-layered systems using
Green function formalism.
The screening of the Coulomb interaction is considered in
the framework of Thomas-Fermi approximation, and a
qualitatively new mechanism of screening by indirect 
excitons is taken into account.
The exciton density is shown to decrease sharply with 
increasing electron-hole separation up to one   
 exciton Born radius.
The strong mutual enhancement of screening and charge-separation
effects is found.\\

\end{abstract}

\section{Introduction} 
Spatially separated quasi-two-dimensional (2D)
electron-hole plasma can be found in different modern 
nano-heterostructure-based devices.
There is a continuous attention to closely situated (so called 
double) quantum wells. This attention is mostly due to the 
possibility of the Bose-Einstein condensation of spatially 
indirect excitons in such a system\cite{Lozovik}. 
Another physical object with spatially separated 
electron-hole plasma, which currently attracts increasing 
interest, is a GaN/AlGaN quantum well, grown on a
sapphire substrate\cite{Kavokin}. 
In the latter case electrons and holes are separated by huge 
(of the order of one MV/cm) internal electric field.
Because of the enhanced binding energy in wide-gap semiconductors,
excitons in nitride-based nanostructures are a subject of
extensive experimental and theoretical research.  

The aim of this work is to provide a consistent many-body 
theory, describing statistical mechanics of the quasi-2D 
exciton/free carrier plasma with spatial charge separation. 

\section{Screening in the Spatially-Separated Electron-Hole System}
First of all, the potentials of electron-electron and electron-hole
interaction should be established.
The Thomas-Fermi (or, as it is often called in the 2D case, Stern-Howard) 
approximation is one of the most frequently used methods for dealing 
with static screening of Coulomb interaction in semiconductor 
systems\cite{Ando}.   
In the frame of this approximation the charge, 
induced by external perturbation, 
originates from a redistribution of free carriers and 
is proportional to a local electric field. 
The resulting electric field potential is given by the Poisson 
equation in the form:
\begin{equation}
\Delta\phi-2q_s^e\bar{\phi_e} g_e(z)-2q_s^h\bar{\phi_h} g_h(z)=
-\frac{4\pi}{\kappa} Q_{ext},
\label{EP2}
\end{equation}
where $\bar{\phi_{a}}(\rho)=\int\phi(\rho,z)g_a(z)\,dz$, 
$g_a(z)$ is a free-carrier-distribution function in the direction
normal to the quantum well plane,
$\kappa$ is a background dielectric constant, $Q_{ext}$ is an external
charge density, and the screening parameter in the 2D case is 
\begin{equation}
q_s^a=
\frac{2\pi e^2}{\kappa}\frac{\partial n_a}{\partial\mu_a}=
\frac{2}{a_B}\frac{m_a}{m_{eh}}\left[
1-\exp\left(-\beta{\pi\hbar^2 n_a}/{m_a}\right)\right].
\label{qs}
\end{equation}
Here $n_a$ is a free particle density, $\mu_a$ is a quasi-Fermi level,
$m_a$ is an effective mass, $\mu_{eh}=m_em_h/(m_e+m_h)$ is the reduced
effective mass, $a_B=\kappa\hbar^2/(\mu^*e^2)$ is the 3D exciton Bohr
radius, and $\beta=1/(k_BT)$.  

The contribution of excitons to screening in the conventional single
quantum well can be treated within the Thomas-Fermi approximation as a 
reduction of free carrier densities entering Eq.(\ref{qs}).    
In the case of spatially separated plasma, screening by excitons
can not be neglected even in the Thomas-Fermi approximation.
The indirect exciton has the finite dipole moment, and the external
electric field causes redistribution of excitons, which 
plays its own role in the resulting field reduction. 
Using the same approach as for a free-carrier plasma, the 
indirect-exciton part of the induced charge density can be written as
\begin{equation}
Q_{ind}^{exc}=
\frac{\kappa}{2\pi}q_s^{exc}(g_h(z)-g_e(z))
\left[e\bar{\phi_{h}}(\rho)-e\bar{\phi_{e}}(\rho)\right].
\end{equation}
Here we introduced the excitonic screening parameter
\begin{equation} 
q_s^{exc}={2\pi e^2 \over \kappa}\frac{\partial n_{exc}}{\partial\mu},
\label{qsexc}
\end{equation}
where $n_{exc}$ is the exciton density and $\mu=\mu_e+\mu_h$.

In the present paper we consider a simple model
of spatially separated electron-hole plasma, where electrons  
and holes are purely two-dimensional and reside in different planes
with distance $d$ between them. The generalization of this model 
to incorporate finite-width quantum wells will be presented somewhere else.
The two-plane model is particularly valuable for studying the interplay 
between screening and spatial separation of different types of carriers. 
The problem of finding electron-electron repulsion and electron-hole
attraction potentials has cylindrical symmetry, so we use
the following representation:
\begin{equation}
\phi(\rho,z)=\frac{1}{2\pi}\int_0^{\infty}\phi_q(z)J_0(\rho q)q\,dq,
\end{equation}   
where $J_0$ is the Bessel function. 
Let us assume that the electronic charge resides in the plane with
coordinate $z=d$ and the holes are in the plane $z=0$. 
In this model the distribution functions $g_a(z)$ are given by 
delta-functions.
The Poisson equation is reduced to a one-dimensional problem:
$$
\frac{d^2}{dz^2}\phi_q(z)-q^2\phi_q(z)-
2q_s^h\phi_q(0)\delta(z)-2q_s^e\phi_q(d)\delta(z-d)
$$
\begin{equation}
-2q_s^{exc}(\delta(z)-\delta(z-d))[\phi_q(0)-\phi_q(d)]
=-\frac{4\pi e}{\kappa}\delta(z-d)
\label{EP3}
\end{equation}
Solving Eq.(\ref{EP3}), we obtain repulsion and attraction 
potentials in $q$-representation: 
\begin{equation}
V_s^{eh}(q)=e\phi_q(0)=
-\frac{2\pi e^2}{\kappa}\frac{qe^{-qd}+q_s^{exc}(1-e^{-2qd})}
{\Theta+q_s^{exc}\Xi},
\label{veh2}
\end{equation}
\begin{equation}
V_s^{ee}(q)=-e\phi_q(d)=
\frac{2\pi e^2}{\kappa}\frac{q+(q_s^h+q_s^{exc})(1-e^{-2qd})}
{\Theta+q_s^{exc}\Xi},
\label{vee2}
\end{equation}
where both terms
$
\Xi=[1-\exp(-qd)]
[2k+(q_s^e+q_s^h)(1+\exp(-qd))],
$
and
$
\Theta=(q+q_s^{(e)})(q+q_s^{(h)})-q_s^{(e)}q_s^{(h)}
\exp(-2qd)
$ 
in denominators of Eqs.(\ref{veh2},\ref{vee2}) do not depend on the 
excitonic screening parameter $q_s^{exc}$. 

\section{Correlated density}
In order to provide a consistent treatment of many-body effects in a
2D electron-hole plasma, we base our theory on a quasi-particle picture
and Green functions technique. Electrons and holes are redistributed
among the quasi-particle states with a sharp spectral peak and  
dispersion $\epsilon_a=\hbar^2 k^2/2m_a$ 
(where the energy is measured from the corresponding size-quantized
level renormalised by the many-body shift) 
and some other "correlated" states originating from the 
interaction between quasi-particles. 

The total density of electrons can be divided into two parts, 
the first part representing ``free'' particles with a quasi-Fermi 
level $\mu_e$, and the second part incorporating interaction between
quasiparticles. This second part, which is called the correlated density, 
is given\cite{ZimmermannAndSchtolz85} by  
\begin{equation} 
n_e^{corr}=2\sum_k \int\frac{d\hbar\omega}{\pi}\Gamma_e(k,\omega)
\left[
f_e(\hbar\omega)-f_e(\epsilon_e(k))
\right]
\frac{\partial}{\partial\hbar\omega}
\frac{\rm P}{\epsilon_e(k)-\hbar\omega}.
\label{korplot}
\end{equation}
Here $\Gamma_e(k,\omega)={\rm Im}\Sigma_e(k,\omega)$
is an imaginary part of the single-particle self-energy, 
$f_a(\epsilon)=[\exp(\beta(\epsilon-\mu_a))+1]^{-1}$ is the Fermi 
distribution function, and the derivative of the principal value
is given
$$
\frac{\partial}{\partial\hbar\omega}
\frac{\rm P}{\epsilon-\hbar\omega}=\lim_{\nu\rightarrow 0}
\frac{(\epsilon-\hbar\omega)^2-\nu^2}
{[(\epsilon-\hbar\omega)^2+\nu^2]^2}.
$$ 
From the point of view of diagrammatic technique, 
exciton is a succession of electron-hole scatterings,
which is represented by the series of diagrams
serving as a basis for the ladder approximation\cite{Mahan}.  
We use the ladder approximation to our many-body problem, 
keeping only ladder-type diagrams in the
expression for the electron self energy:
\begin{equation}
\Sigma_e(\vec{k},\Omega_e)=-\frac{1}{\beta}\sum_{b,\vec{k}',z_b}
\left[2\tilde{T}_{eb}(\vec{k},\vec{k}-\vec{k}',\vec{k},\Omega_e+\Omega_b)-
\delta_{eb}\tilde{T}_{eb}(-\vec{k}',\vec{k}-\vec{k}',\vec{k}, \Omega_e+
\Omega_b)
\right] G_b(-\vec{k}, \Omega_b),
\end{equation}
where the summation is taken over different types of particles, $b=e,h$,
2D wave-vector $\vec{k}'$ and Fermi-type Matsubara frequencies $\Omega_b$.
The quantity $\tilde{T}$ represents a sum of ladder diagrams 
for electron-electron or electron-hole scattering and is given by 
the frequency-dependent T-matrix equation:  
$$
\tilde{T}_{ab}(\vec{k},\vec{q},\vec{k}'', \Omega)=
$$
\begin{equation}
-V_s^{ab}(\vec{k}-\vec{k}'')
+\sum_{\vec{k}'} V_s^{ab}(\vec{k}-\vec{k}')
G_{ab}(\vec{k}',\vec{q}, \Omega)
\tilde{T}_{ab}(\vec{k}',\vec{q},\vec{k}'',\Omega),
\label{TE}
\end{equation}
where the potentials $V_s^{ab}$ are given by Eqs.(\ref{veh2},\ref{vee2}),
$$
G_{ab}(\vec{k},\vec{q}, \Omega)=
\frac{N_{ab}(\vec{k},\vec{q})}
{\epsilon_a(\vec{k})+\epsilon_b(\vec{q}-\vec{k})-\hbar\Omega},
$$ 
and
$$
N_{ab}(\vec{k},\vec{q})=
1-f_a(\epsilon_a(\vec{k}))-f_b(\epsilon_b(\vec{q}-\vec{k})).
$$    
Following Zimmermann \cite{Zimmermann}, who derived a similar formula
for the 3D case, we obtain the expression for 
the correlated density in the 2D electron-hole system:
\begin{equation}
n_a^{corr}=\frac{2}{\beta\pi\hbar^2}\sum_{m=-\infty}^{+\infty}
\left[
\sum_n M_{eh}L_{eh}
\left(\epsilon_{mn}\right)
+
\frac{1}{\pi}\sum_{b=e,h}M_{ab}\lambda^{ab}_m\int_0^{\infty}dk
L_{ab}
\left(\frac{\hbar^2k^2}{2m_{ab}}
\right)2\sin^2\delta_m^{ab}(k)
\frac{d\delta_m^{ab}}{dk}
\right],
\label{OMAL}
\end{equation}
$$
L_{ab}(\epsilon)=
-\ln\left[
1-\exp\left(\beta({\mu_a+\mu_b-\epsilon})\right)
\right],
$$
$$
M_{ab}=m_a+m_b,~~~~~~~ m_{ab}=\frac{m_am_b}{m_a+m_b}, ~~~~~~
\lambda^{ab}_m=1-\delta_{ab}(-1)^m/2.
$$
Here $\epsilon_{mn}$ are the values of the exciton bound state energies,
where the first index $m$ denotes the angular momentum and the 
second index $n$ numerates different states for a given $m$.
The bound state energies correspond to the poles of the T-matrix. 
The quantities $\delta_m^{ab}(k)$ give the phases of the coefficients 
in the Fourier series expansion of the T-matrix. 
The values of these quantities coincide with the scattering phase shifts 
when the k-space filling is neglected. 
Thus, the second part of Eq.(\ref{OMAL}) describes the scattering
states contribution to the correlated density.
In the low density/high temperature limit this expression differs
from that of Portnoi and Galbright \cite{Portnoi} only by the
 factor $2\sin^2\delta_m^{ab}$ in the integrand.
The ``e-e'' part of the correlated density has a negative sign 
and is responsible for the reduction of the plasma density 
due to repulsion between the carriers of the same type.
The electron-hole part of the density is positive and 
can be attributed to the presence of the excitons.
As it was shown in Ref.\cite{Portnoi} for dilute plasma,
this part of the density does not change abruptly
when one of the bound states disappears with increasing screening. 
We treat the electron-hole part of the correlated density as the 
density of excitons:
\begin{equation}
n_{exc}=
\frac{2}{\beta}\frac{M_{eh}}{\pi\hbar^2}\sum_{m=-\infty}^{+\infty}
\left[
\sum_n L_{eh}
\left(\epsilon_{mn}\right)
+\frac{1}{\pi}
\int_0^{\infty}dk
L_{eh}
\left(\frac{\hbar^2k^2}{2m_{eh}}
\right)
2\sin^2\delta_m^{eh}(k)
\frac{d\delta_m^{eh}}{dk}
\right].
\label{nexc}
\end{equation}  
Substituting $n_{exc}$ from Eq.(\ref{nexc}) into Eq.(\ref{qsexc}) we
get the following expression for the excitonic screening parameter: 
\begin{equation}
q_{exc}=
\frac{4M_{eh}e^2}{\kappa\hbar^2}\sum_{m=-\infty}^{+\infty}
\left[
\sum_n f_B\left(\epsilon_{mn}\right)
+
\frac{1}{\pi}\int_0^{\infty}dk
f_B
\left(\frac{\hbar^2k^2}{2m_{eh}}
\right)
2\sin^2\delta_m^{eh}(k)
\frac{d\delta_m^{eh}}{dk}
\right],
\label{qsE}
\end{equation}
$$
f_B(\epsilon)=
\left[
\exp(\beta(\epsilon-\mu_e-\mu_h))-1
\right]^{-1}.
$$
To calculate the exciton density we have to run the following 
self-consistent procedure. Firstly, for given electron-hole 
density we calculate the screening parameters from Eq.(\ref{qs})
and obtain corresponding bound state energies and phase shifts. 
Then, using these results we calculate the new values of excitonic 
screening parameter and free carrier density and recalculate
bound state energies and phase shifts, repeating this 
procedure until the exciton density stops changing.

\section{Results and discussion} 
Throughout the calculations we neglect the phase shift
dependence on the carrier density and obtain the
values of $\delta_{ab}$ using the 2D modification\cite{Portnoi} 
of the variable phase approach. 
To find the bound state energies we use
the method, proposed by Campi and co-workers\cite{Campi}, which
is also based on the ladder approximation. We take into account only
the ground exciton state. The material parameters are chosen
to match these of the GaAs/AlGaAs quantum wells, and the difference
between dielectric constants of the QW and the barrier is neglected.
 
In Figure 1 we present the dependence of the binding energy of the
exciton ground state on the inter-plane distance. 
One can see (Fig. 1b) that the decrease of the exciton binding energy 
with increasing the distance is substantially enhanced by the presence 
of free carriers and indirect excitons, which screen the Coulomb attraction. 
Figure 2 shows the exciton density as a function of the total plasma 
density for different inter-plane distances.
One can see that the exciton density has a maximum and then
decreases sharply  with increasing the total number of carriers,
which reflects the action of both screening and k-space filing.

\newpage

\begin{figure}
\begin{center}
\includegraphics{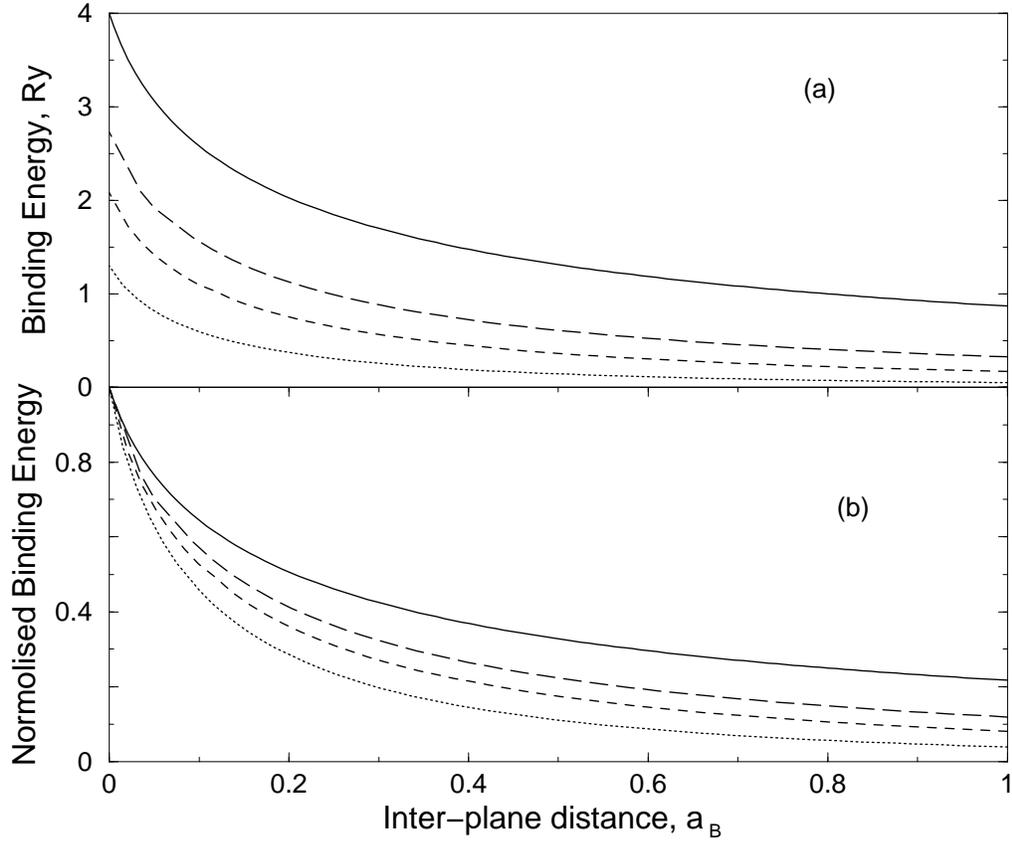}
\end{center}
\vskip 14truecm
\caption{
The ground state binding energy of the indirect exciton
as a function of inter-plane separation for different
free-carrier screening parameter: $q_s^e=0$ (solid line), 
$q_s^e=0.25$ $a_B^{-1}$ (long-dashed line), $q_s^e=0.5$ $a_B^{-1}$ 
(dashed line), 
$q_s^e=a_B^{-1}$ (dotted line). 
(a) - measured in the 3D exciton Rydberg units, 
(b) - normalized by its maximum value. 
}
\end{figure}

\newpage

\begin{figure}
\begin{center}
\includegraphics{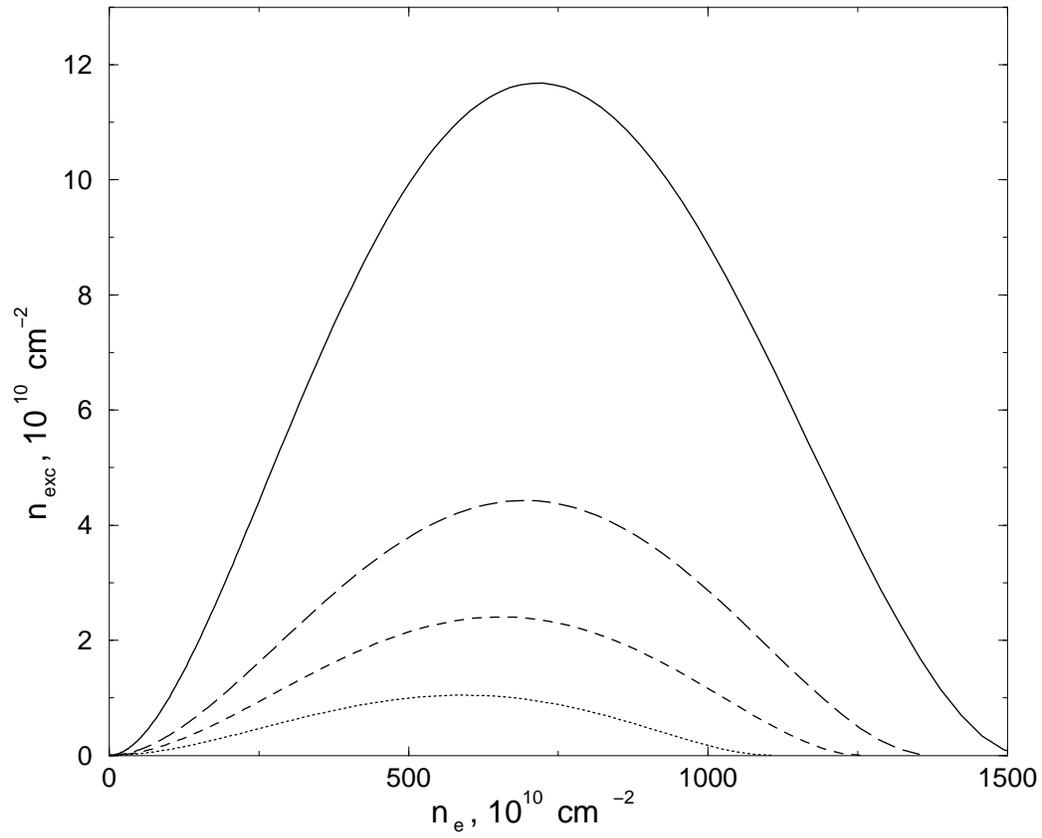}
\end{center}
\vskip 14truecm
\caption{
Exciton density as a function of plasma density for different
inter-plane distances: $d=0$ (solid line), $d=0.25$ $a_B$ (long-dashed line),
$d=0.5$ $a_B$ (dashed line),  $d=a_B$ (dotted line). 
}
\end{figure}
\end{document}